\documentclass[aps,superscriptaddress,showpacs,floatfix]{revtex4}
\usepackage{amssymb}
\usepackage{amsmath}
\usepackage{graphicx}
\usepackage[normalem]{ulem}
\usepackage[dvips]{color}
\usepackage{multirow}
\usepackage[bookmarksnumbered,bookmarksopen,colorlinks,citecolor=blue,linkcolor=blue]{hyperref} %

\usepackage{CJK}                     
\usepackage{bm}                      
\usepackage{mathptmx}                
\usepackage{dcolumn}                 

\begin{document}


\title{Magnetic effects in heavy-ion collisions at intermediate energies}

\author{Li Ou}
 \email{only.ouli@gmail.com}
 \affiliation{Department of Physics and Astronomy, Texas A\&M University-Commerce, Commerce, TX 75429-3011, USA}
 \affiliation{College of Physics and Technology, Guangxi Normal University, Guilin, 541004, P. R. China}

\author{Bao-An Li\footnote{Corresponding author. Bao-An\_Li@tamu-commerce.edu}}
 \affiliation{Department of Physics and Astronomy, Texas A\&M University-Commerce, Commerce, TX 75429-3011, USA}

\date{\today}

\begin{abstract}
The time-evolution and space-distribution of internal
electromagnetic fields in heavy-ion reactions at beam energies
between 200 and 2000 MeV/nucleon are studied within an
Isospin-dependent Boltzmann-Uhling-Uhlenbeck transport model IBUU11.
While the magnetic field can reach about $7\times 10^{16}$ G which
is significantly higher than the estimated surface magnetic field ($\sim 10^{15}$
G) of magnetars, it has almost no effect on nucleon observables as
the Lorentz force is normally much weaker than the nuclear force.
Very interestingly, however, the magnetic field generated by the
projectile-like (target-like) spectator has a strong
focusing/diverging effect on positive/negative pions at
forward (backward) rapidities. Consequently, the differential
$\pi^-/\pi^+$ ratio as a function of rapidity is significantly
altered by the magnetic field while the total multiplicities of both
positive and negative pions remain about the same. At beam energies
above about 1 GeV/nucleon, while the integrated ratio of total
$\pi^-$ to $\pi^+$ multiplicities is not, the differential
$\pi^-/\pi^+$ ratio is sensitive to the density dependence of
nuclear symmetry energy $E_{\rm{sym}}(\rho)$. Our findings suggest
that magnetic effects should be carefully considered in future
studies of using the differential $\pi^-/\pi^+$ ratio as a probe of
the $E_{\rm{sym}}(\rho)$ at supra-saturation densities.

\end{abstract}

\maketitle

\section{Introduction}
Magnetic fields exist everywhere in the Universe. To set the scale
and appreciate the strong magnetic fields created during heavy-ion
collisions, we first recall the magnitudes of several typical
magnetic fields from various sources. Many spiral galaxies have
magnetic fields with a typical strength of $\sim 3\times 10^{-6}$ G
\cite{Sofue86} and it is estimated that the intergalactic magnetic
fields presently have an intensity of about $\leq 10^{-9}$ G
\cite{Kawabata69}. Some people believe that the present magnetic
field of the Universe is amplified from a seed about $10^{-20}$ G by
the dynamo mechanism \cite{Dimopoulosa97,Grasso01} while magnetic
fields up to $10^{24}$ G might appear in the early Universe
\cite{Grasso01}. The strongest magnetic field of about $10^{15}$ G
near the surfaces of magnetars \cite{Kouveliotou98,Ibrahim02} or
even higher ($10^{16}$-10$^{17}$ G) associated with the cosmological
gamma-ray bursts \cite{Ruderman00} have been found from
astrophysical observations. Due to the limit of tensile strength of
terrestrial materials, the strongest man-made steady magnetic field
is only about $4.5 \times 10^{5}$ G. To our best knowledge, it was
first pointed out by Rafelski and M$\rm{\ddot{u}}$ller that, in
addition to strong electrical fields, unusually strong magnetic fields
are also created in heavy-ions collisions (HICs). In sub-Coulomb
barrier U+U collisions, the magnetic field was estimated to be on
the order of $10^{14}$ G \cite{Rafelski76}. More recently, it has
been shown by Kharzeev et al. that HICs at RHIC and LHC can create
the strongest magnetic field ever achieved in a terrestrial
laboratory \cite{Kharzeev08}. For example, in noncentral Au+Au
collisions at 100 GeV/nucleon, the maximal magnetic
field can reach about $10^{17}$ 
G \cite{Kharzeev08,Skokov09}. It thus provides a unique environment
to investigate the Quantum Chromodynamics (QCD) at the limit of high
magnetic field. Indeed, the study of quark-gluon-plasma under strong
magnetic field has attracted much attention by the high energy
heavy-ion community, see, e.g., ref. \cite{Kha11} and references
therein. In particular, it has been shown theoretically that
\cite{Kharzeev06,Kharzeev07,Kharzeev08,Voloshin10} QCD topological
effects in the presence of very intense electromagnetic fields,
i.e., the ``Chiral Magnetic Effect'', may be an evidence of local
parity violation in strong interactions. Experimentally, interesting
indications have been reported, see, e.g., refs.
\cite{Abelev09,Abelev10}.

Stimulated by the interesting findings at RHIC and realizing that
all transport model studies of magnetic effects have so far focused
on high energy HICs \cite{Skokov09,Voronyuk11}, we investigate in
this work first the strength, duration and distribution of internal
magnetic fields created in HICs at beam energies between 200 and
2000 MeV/nucleon. This is the beam energy range covered by several
accelerators in the world. We then focus on identifying possible
magnetic effects on experimental observables using an
isospin-dependent Boltzmann-Uhling-Uhlenbeck (BUU) transport model
IBUU11 \cite{BALi04,Li11}. We find that while the magnetic field can
reach about $7\times 10^{16}$ G in these reactions, it has almost no
effect on nucleon observables as the Lorentz force is negligibly
small compared to the nuclear force. Very interestingly, however,
the magnetic field generated by the projectile-like (target-like)
spectator moving forward (backward) in the center of mass frame has
a strong focusing/diverging effect on positive/negative pions moving
forward (backward). As a result, the differential $\pi^-/\pi^+$
ratio as a function of rapidity is significantly altered by the
magnetic field while the total $\pi^-$ and $\pi^+$ multiplicities
remain about the same.

The paper is organized as follows. In the next section, we outline
how the internal electromagnetic fields in HICs are calculated in
the IBUU11 transport model. The characteristics of the
electromagnetic fields and their effects on several experimental observables in intermediate energy HICs
are then discussed in Section III. Finally, a summary is given at
the end.

\section{The model}
In the presence of electrical and magnetic fields $\bm{E}$ and $\bm{B}$,
the BUU equation can be written as
\begin{eqnarray}\label{BUU}
\left[\frac{\partial}{\partial
t}+\frac{\bm{P}}{E}\nabla_r-(\nabla_r\bm{U}-q\bm{v}\times
\bm{B}-q\bm{E})\nabla_p\right]f(\bm{r},\bm{p},t)
=I(\bm{r},\bm{p},t)
\end{eqnarray}
where $I(\bm{r},\bm{p},t)$ is the collision integral  simulated by
using the Monte Carlo method. The electrical field $\bm{E}$ (Coulomb
field) has already been considered in most transport models. To
include consistently both the electrical and magnetic fields
satisfying Maxwell's equations, the Li$\rm{\acute{e}}$nard-Wiechert
potentials at a position $\bm{r}$ and time $t$ are evaluated
according to
\begin{eqnarray}\label{Ef}
e \bm{E}(\bm{r},t) &=&\frac{e^2}{4\pi \epsilon_0}\sum_n
Z_n\frac{c^2-v_n^2} {(c R_n-\bm{R}_n \cdot \bm{v}_n )^3  }
(c\bm{R}_n - R_n\bm{v}_n)
\end{eqnarray}
and
\begin{eqnarray}\label{Bf}
 e \bm{B}(\bm{r},t) &=&\frac{e^2}{4\pi \epsilon_0 c}\sum_n Z_n\frac{c^2-v_n^2}
{(c R_n-\bm{R}_n \cdot \bm{v}_n )^3  } \bm{v}_n \times \bm{R}_n
\end{eqnarray}
where $Z_n$ is the charge number of the $n$th particle.
$\bm{R}_n=\bm{r}-\bm{r}'_n$ is the relative position of the field
point $\bm{r}$ with respect to  the position $\bm{r}'_n$ of particle
$n$ moving with velocity $\bm{v}_n$ at the retarded time $t_{{\rm
r}n}=t-|\bm{r}-\bm{r}'_n(t_{{\rm r} n})|/c$. The summation runs over
all charged particles in the reaction system. In non-relativistic
cases, i.e., all particles satisfy the condition $v\ll c$, the Eq.
\eqref{Ef} and \eqref{Bf} reduce to the classical expressions
\begin{eqnarray}\label{Efc}
e \bm{E}(\bm{r},t) &=&\frac{e^2}{4\pi \epsilon_0}\sum_n Z_n\frac{1}
{R_n^3  } \bm{R}_n
\end{eqnarray}
and
\begin{eqnarray}\label{Bfc}
 e \bm{B}(\bm{r},t) &=&\frac{e^2}{4\pi \epsilon_0 c^2}\sum_n Z_n\frac{1} {R_n^3  } \bm{v}_n \times
 \bm{R}_n.
\end{eqnarray}
The first equation is essentially the Coulomb's law, and the latter is the
Bio-Savart law for a system of moving charges.

To take into account accurately the retardation effects, the phase
space information of all nucleons before the moment $t$ are required to
calculate the electromagnetic fields at that moment. Some special
care is thus necessary in initializing the reaction. In principle, the
two colliding nuclei should be initialized to come from infinitely
far away towards each other on their Coulomb trajectories. In
practice, considering the need of keeping the initial nuclei stable
and the computing time low, the initial distance between the
surfaces of the two colliding nuclei is taken as 3 fm in our
calculations. We make a pre-collision phase space history for all
nucleons assuming that they are frozen in the projectile/target
moving with a center of mass velocity $\bm{v}_{p/t}$, i.e.,
$\bm{r}_i=\bm{r}_i^0+\bm{v}_{p/t}\cdot t$, where $\bm{r}_i^0$ is the
initial coordinate of the nucleon. As we shall show, comparisons of our transport
model calculations with analytical estimates for two moving charges
(target and projectile) in both relativistic and non-relativistic
cases indicate that our method of handing the pre-collision
phase-space histories of all nucleons is reasonable.

\begin{figure}[htpb]
\includegraphics[width=0.5\textwidth]{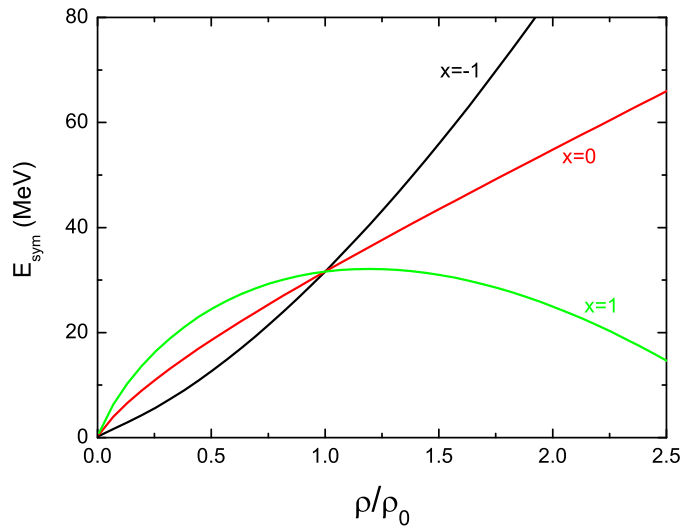}
    \caption{Density dependence of nuclear symmetry energy used in the IBUU11 calculations.}
    \label{SymE}
\end{figure}
We refer the BUU code used in this study IBUU11.  Compared to the
IBUU04 \cite{BALi04} where the MDI (Momentum-Dependent-Interaction)
is used \cite{Das03}, besides the electromagnetic fields with
retardation effects, an isospin-dependent three-body force
\cite{Xu10b} (instead of the standard one used in the MDI, Gogny and
Skyrme effective interactions) is used. Moreover, the high-momentum
tail of the MDI isoscalar potential is readjusted to better fit the
nucleon optical potential from nucleon-nucleus scattering
experiments. Details of these modifications and their effects on
experimental observables will be presented in a forthcoming
publication \cite{Li11}. In this work, we focus on the magnetic
aspect of HICs at intermediate energies. Since one of our main
motivations here is to see whether experimental observables known to
be sensitive to the $E_{\rm{sym}}(\rho)$ is affected by the magnetic
effects, we notice here that in the IBUU11 the $E_{\rm{sym}}(\rho)$ is
controlled by a parameter $x$ introduced in the three-body part of
the MDI interaction \cite{Das03,Xu10b}. By adjusting the parameter $x$ one
can mimic diverse behaviors of the $E_{\rm{sym}}(\rho)$ predicted by
various microscopic many-body theories \cite{LCK08}. As an example,
shown in Fig. \ref{SymE} are the $E_{\rm{sym}}(\rho)$ with $x=1, 0$
and $-1$, respectively.

\section{Results and discussions}
In this section, we first illustrate and discuss the beam energy and
impact parameter dependence of the time-evolution and
space-distribution of magnetic field. To help understand the
magnetic effect in HICs, we shall also compare the Lorentz force
with the Coulomb and nuclear forces. We then present and discuss
magnetic effects on experimental observables.

\subsection{Characteristics of internal electromagnetic fields in heavy-ion reactions}
\begin{figure}[htpb]
\includegraphics[width=0.5\textwidth]{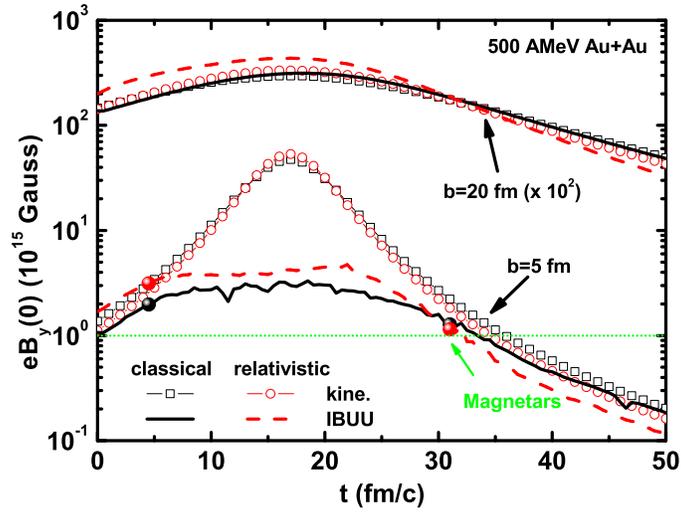}
\caption{(color online) Time evolutions of the magnetic field
strength $eB_y(0)$ at the center of mass of the reaction system for
500 AMeV Au+Au reactions at $b$=5 and 20 fm, respectively. The
magnitude of $eB_y(0)$ for $b$=20 fm is multiplied by $10^2$ for
clarity. The approximate beginning and ending of the overlap phase
between the projectile and target are indicated by the small balls
(for $b=$20 fm there is no overlap).}{\label{bb0520}}
\end{figure}

Features of the internal electromagnetic fields are independent of
the symmetry energy parameter $x$.  In this subsection, unless
otherwise specified a value of $x=1/3$ is used. We take the $z$
($x$) axis as the beam (impact parameter) direction. Based on the
formula of magnetic field strength in Eq. \eqref{Bf}, the dominant
component of the internal magnetic field is in the $y$ axis
perpendicular to the reaction plane ($z-x$). The component in the
reaction plane is negligible because of the slow motions of nucleons
in the $x$ or $y$ directions especially in the early phase of the
reaction. To test our approach used in calculating the
electromagnetic fields, we first compare the magnetic field $B_y(0)$
at the center of mass of the reaction system calculated using the
full IBUU11 dynamically with those obtained under some limiting
conditions for idealized situations. Shown in Fig. \ref{bb0520} are
the values of $B_y(0)$ for Au+Au reactions at a beam energy of 500
AMeV and an impact parameter of $b$=5 and 20 fm, respectively. As a
reference, the approximate magnetic field of $10^{15}$ G on the
surfaces of magnetars is also indicated. The legend ``classical''
and ``relativistic'' indicate results obtained using Eq. \eqref{Bfc}
and Eq. \eqref{Bf}, respectively. For a comparison, we have also
performed calculations using both Eq. \eqref{Bfc} and Eq. \eqref{Bf}
assuming that the projectile and target are two point charges
located at their individual centers of masses and are moving with
their initial velocities only. Results of this calculation are
denoted by the ``kine.''.
Several interesting observations can be made. Firstly, it is seen
that the $B_y(0)$ calculated with the classical and relativistic
formulas are very close to each other, for both the kinematic and
dynamical calculations, as one expects for reactions at relatively
low beam energies. Secondly, the dynamical IBUU11 results and the
kinematic estimates are very close at the beginning and the end of
the reaction, but they are very different during the reaction phase
spanned by the small balls of the same color. The magnetic field has
contributions from the projectile-like and target-like spectators as
well as charged particles in the participant region. Contributions
from the latter, however, are very weak because of the approximately
isotropic nucleon momentum distribution there. Once the projectile
and target begins overlapping, nucleon-nucleon collisions will start
transferring the participants' longitudinal momenta into transverse
directions. Thus, the $B_y(0)$ from the IBUU11 is weaker than the
kinematic estimate during the reaction phase. We notice that the
magnetic field in the $x$ and $z$ directions are rather weak because
they only come from charged participants which are moving
essentially randomly in all possible directions. For the very
peripheral reactions with $b$=20 fm, the two nuclei do not overlap.
As one expects, thus there is almost no difference between the
kinematic and dynamical results. The above comparisons enhance our
confidence in using the IBUU11 model to study the internal
electromagnetic fields and their effects in HICs. In the following,
we only present results calculated with the relativistic formula and
the dynamical IBUU11 model.

\begin{figure}[htpb]
\includegraphics[width=1\textwidth]{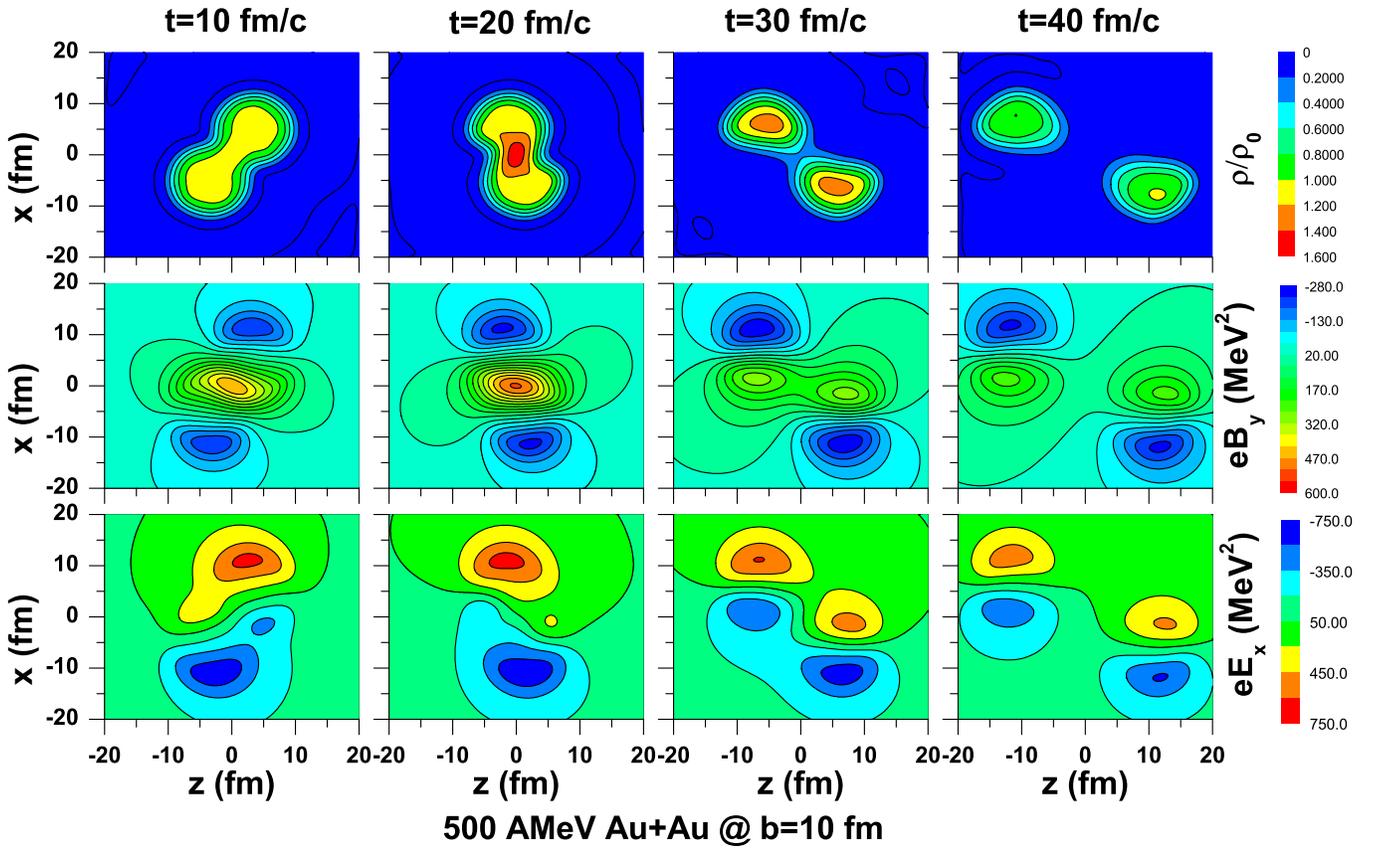}
\caption{(color online) Distributions of the nucleon density
$\rho/\rho_{0}$ (upper panel), the magnetic field strength $eB_y$
(middle panel) and the electrical field strength $eB_x$ (lower panel)
in the $x-z$ plane at $t$=10, 20, 30 and 40 fm/$c$ for the 500 AMeV
Au+Au collisions at an impact parameter of $b$=10 fm.}{\label{fig3D}}
\end{figure}
The contours of the nucleon density $\rho/\rho_{0}$ , the
magnetic field strength $eB_y$, and the electric field strength
$eE_x$ in the $x-z$ plane  at $t$=10, 20, 30 and 40 fm/$c$ for the
500 AMeV Au+Au collisions at an impact parameter of $b$=10 fm are
shown in Fig. \ref{fig3D}. We notice that both the $eB_y$ and $eB_x$
are plotted here in unit of MeV$^2$ which is equal to $1.44\times
10^{13}$ G. For discussing the spatial distribution of the
electromagnetic fields, we can divide the space into three zones in
terms of the $x$ coordinate: the outside-zone where $|x|>$ 15 fm;
the spectator-zone where 5 fm $\leq |x|\leq$ 15 fm; and the
overlap-zone where $|x|<$ 5 fm. As mentioned above, the
electromagnetic fields come from both the spectators and
participants. In the outside-zone, the spectator near the field
point generates a stronger magnetic field in the negative
$y$-direction while the other spectator farther away generates a
weaker magnetic field in the positive $y$-direction. The
superposition leads to a magnetic field points to the negative
$y$-direction. On the other hand, the electric field $eE_x$ in the
outside-zone includes contributions from all charges. Its sign is
the same as the sign of the $x$-coordinate of the field point. In
the overlap-zone, the magnetic fields generated by the two
spectators will superimpose constructively since they are all in the
positive $y$-direction, while the magnetic fields generated there by
the moving charges in the participant region will largely cancel
each other. The strength of the magnetic field peaks when the two
nuclei have reached the maximum compression. It then drops when the
spectators depart from each other. The signs of the electric filed
in the $x$-direction generated by the two spectators are always
opposite, leading to the very weak electrical field in the
participant region where the magnetic field is the strongest.

\begin{figure}[htpb]
\includegraphics[width=1.\textwidth]{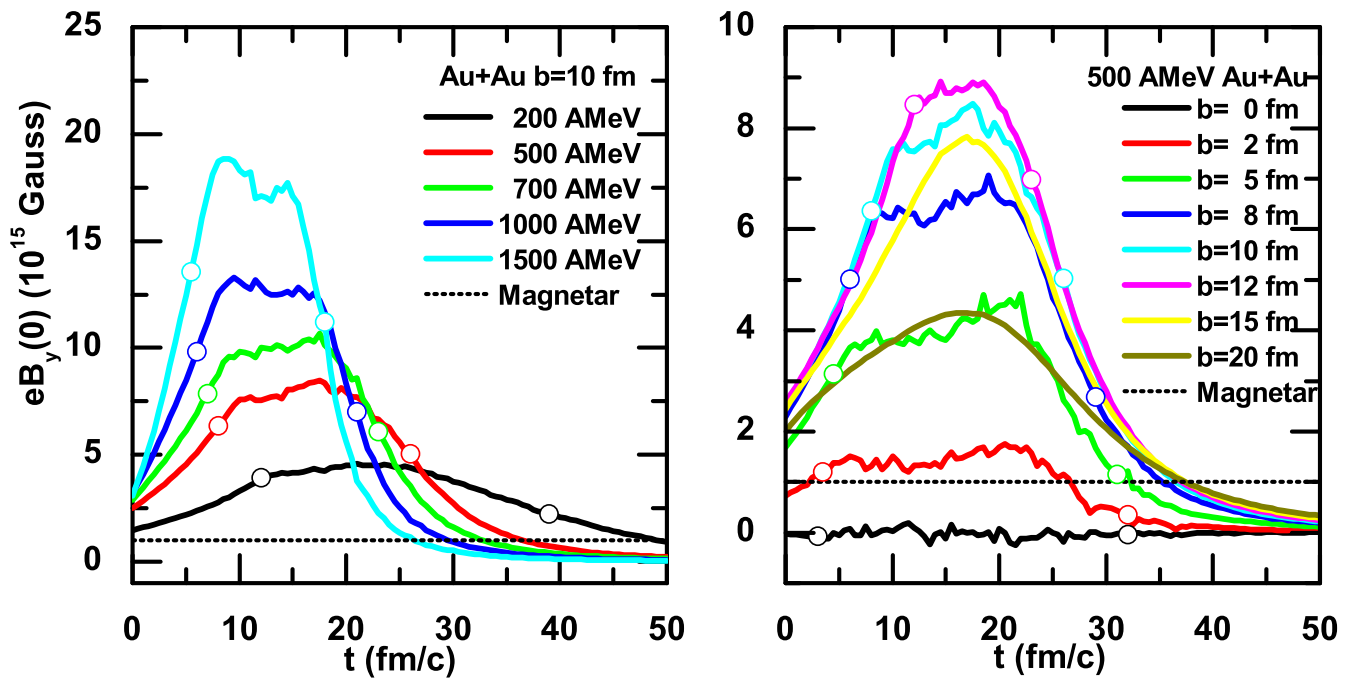}
\caption{(color online) (Right panel) Impact parameter dependence of
$eB_y(0)$ for the 500 AMeV Au+Au collisions. (Left panel) Beam energy dependence of
$eB_y(0)$ for the Au+Au collisions with $b=$10 fm. Durations of the overlap between the projectile and
target are indicated roughly by the open circles (for $b=$15, 20 fm
there is no overlapping). The thin black dashed line is for the magnetic field
strength at the surfaces of magnetars.}{\label{figeb}}
\end{figure}

Next, we explore the impact parameter and beam energy dependence of
the magnetic field at the center of mass of the reaction system.
Shown in the right panel of Fig. \ref{figeb} is the impact parameter
dependence of $eB_y(0)$.
The strength of magnetic field grows with increasing impact
parameter $b$ up to about $b=12$ fm. It then starts decreasing with
larger $b$. This is easily understandable. There are basically two
factors determining the magnetic field strength for a given beam
energy. One is the position vector $\bm R$ from the moving charges to the
field point, and the other one is the charge number of the
spectator $N_s$. Their competition determines the strength of the
magnetic field. For head-on collisions, equivalently there are two
counter currents leading to an almost zero magnetic field at the
center of the reaction. For off-central collisions, as the impact
parameter increases, while the spectators are farther away from the
center they carry more charges. The net result is that the magnetic
field becomes stronger with increasing impact parameter. However, as
the impact parameter becomes larger than the sum of the radius of
the projectile and target, e.g., when $b>12$ fm for the Au+Au
reaction, almost all charges are with the spectators, the magnetic
field is thus only determined by the $\bm R$. Therefore, the reactions with
lager impact parameters create weaker magnetic fields at the center
of the reaction. Based on the IBUU11 results, off-central collisions
with $b=8\sim 10$ fm seem to be the most suitable impact parameter
range to produce the strongest magnetic effect. These reactions
create strong magnetic fields and also enough light charged
particles moving in the magnetic fields to be detected in
experiments. Another factor determining the strength of magnetic
field is the velocity of spectators, i.e., the beam energy of the
reaction. Shown in the left panel of Fig. \ref{figeb} is the beam
energy dependence of $eB_y(0)$. As one expects, while the maximum
strength of the magnetic field increases with beam energy the
duration of the strong magnetic field decreases since the spectators
leave the collision region quickly at higher beam energies. Compared
to reactions at RHIC, the strength of the magnetic field is about 10 times lower but the
reaction lasts about 10 times longer. Since observable effects of any force
depend on not only its strength but also its duration, magnetic effects in HICs at intermediate
energies are thus worth an investigation.

\subsection{Magnetic effects on observables in heavy-ion collisions}
While no chiral magnetic effect is expected in HICs at intermediate
energies, it is still interesting to examine magnetic effects on
hadronic observables. First of all, we would like to mention that
the effects of strong magnetic fields on the Equation of State (EOS)
of cold hadronic and quark matter including the Landau quantization
and the nucleon anomalous magnetic moment in neutron stars have been
studied extensively, see, e.g., refs.
\cite{Bro,Car,Ban,Rab08,Rab11}. It has been shown consistently that
the magnetic effects become significant only for magnetic fields
stronger than about $10^{18}$ G. Moreover, at finite temperature
some of the magnetic effects get mostly washed out \cite{Rab11}.
Since the temperature is high and the maximum strength of the
magnetic field created is still below $10^{18}$ G even at RHIC
energies, it is not necessary to consider effects of the magnetic
field on the nuclear EOS. Instead, we focus directly on magnetic effects due to
the Lorentz force acting on moving charges. In the following, we
examine separately magnetic effects on nucleons and pions.

\subsubsection{Lorentz force compared with the Coulomb and nuclear forces}
\begin{figure}[htpb]
\includegraphics[width=0.45\textwidth]{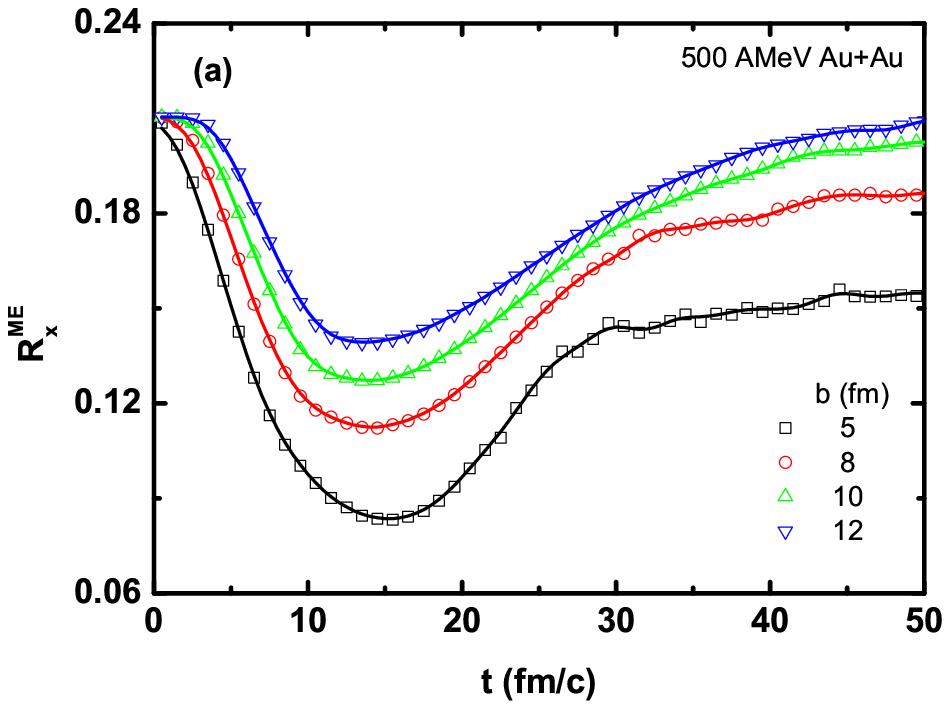}
\includegraphics[width=0.45\textwidth]{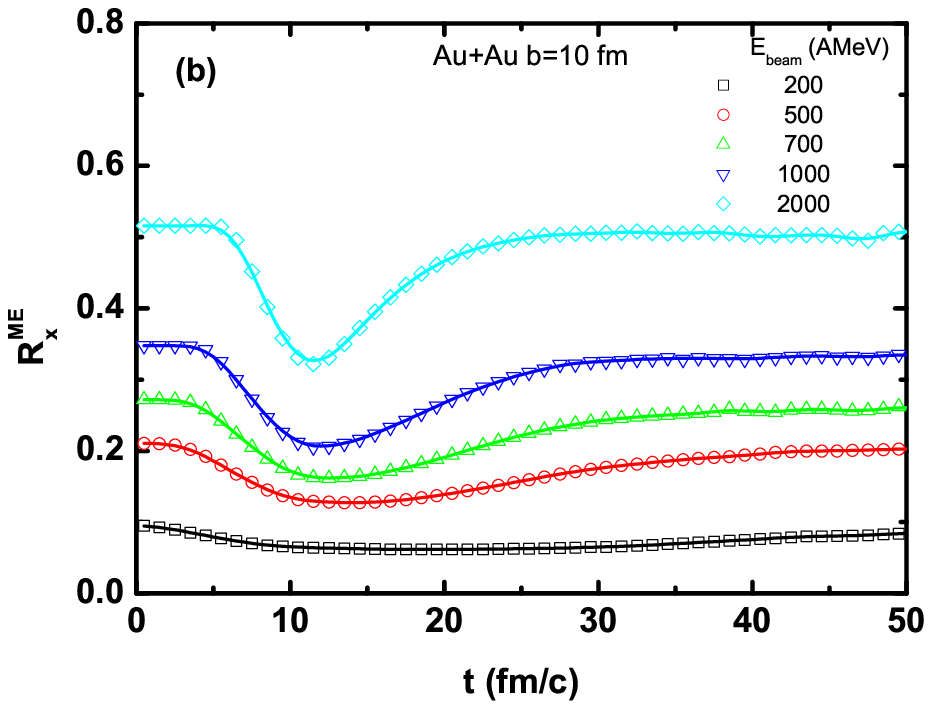}\\
\includegraphics[width=0.45\textwidth]{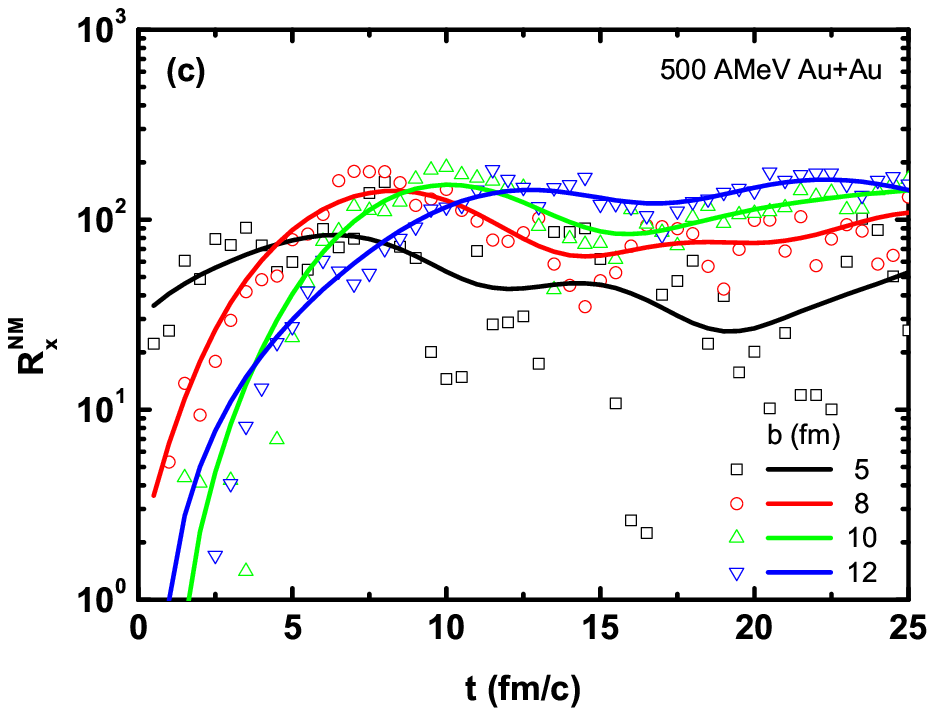}
\includegraphics[width=0.45\textwidth]{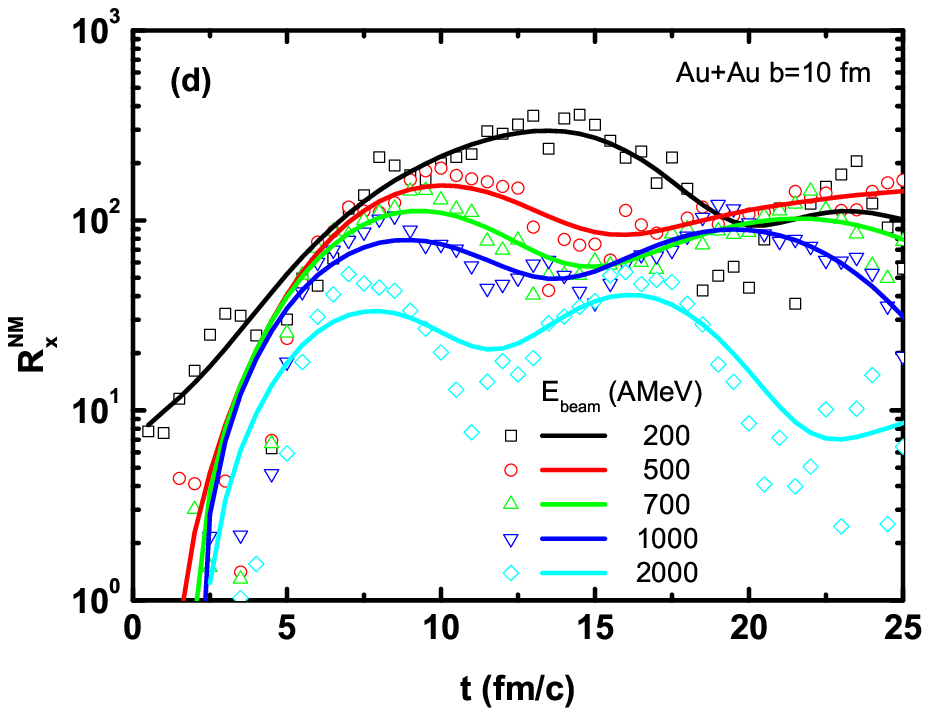}\\
\caption{(color online) The ratio between the $x$-components of the magnetic
and electric forces ($R_x^{ME}$) (a) at various impact
parameters for the 500 AMeV Au+Au reactions, (b) at various incident
energies for the Au+Au reactions with $b$=10 fm;  and the ratio between the
$x$-components of the magnetic and nuclear forces ($R_x^{NM}$) (c) at
various impact parameters for the 500 AMeV Au+Au reactions, (d) at various
incident energies for the Au+Au reactions with $b$=10 fm (the lines are
the results smoothed with the Fast Fourier Transformation Filter to guide the eye), respectively.}{\label{fben}}
\end{figure}
For nucleons, the magnetic effects are expected to be negligible as
the Lorentz force is known to be very small compared to the nuclear
force. On the other hand, while the electrical and magnetic fields
are strongly correlated, the Coulomb force has been routinely taken
into account but the Lorentz force is normally neglected in modeling
HICs. To check the validity of this practice and obtain a more quantitative
understanding about the relative importance of the Lorentz, Coulomb and nuclear forces,
we examine in Fig. \ref{fben} the ratios of the Lorentz force over the Coulomb and
nuclear forces for a test-charge. To be specific, we calculate the
ratio $R_x^{ME}$ of the $x$-component of the Lorentz force over that
of the Coulomb force for a test-charge in the outside-zone. As a
reference, we first make an analytical analysis for a simplified
case. For a test-charge located at the surface of the projectile
moving on the trajectory of $\bm{r}$($-\frac{b}{2}-R$, 0, $z_0+v_{0}
t$), where $R$, $z_0$ and $v_0$ is the radii, initial $z$-coordinate
and the beam velocity, assuming the electromagnetic fields are due
to two moving point charges (projectile and target) given by Eqs.
\eqref{Efc} and \eqref{Bfc}, the $R_x^{ME}$ is simply
\begin{eqnarray}\label{rx}
R_x^{ME}=\frac{F^M_x}{F^E_x}=\frac{e v_zB_y }{e
E_x}=\left(\frac{v_0}{c}\right)^2.
\end{eqnarray}
Thus, it is clear that only for fast moving particles likely
existing in reactions at high beam energies, the Lorentz force is
expected to be significant compared to the Coulomb force. We now
examine numerically the $R_x^{ME}$ for the test-charge using the
electromagnetic fields calculated with the IBUU11. In window (a),
the time evolution of $R_x^{ME}$ is shown for several impact
parameters for the 500 AMeV Au+Au reactions. The evolution can be
approximately divided into four periods. Before the two nuclei get
in touch, $R_x^{ME}=$0.21 which is exactly the same as the
prediction of Eq. \ref{rx}. In the compression phase, since the
magnetic fields in the outside region generated by the
projectile-like and target-like spectators are in the opposite
directions, the net magnetic field decreases whereas the electric
field there becomes stronger. Consequently, the $R_x^{ME}$ drops
until about 15 fm/c. In the expansion phase, the situation is
reversed.  After the collisions are over, the $R_x^{ME}$ keeps
approximately a constant value smaller than $(v_0/c)^2$ depending on
the impact parameter. The beam energy dependence shown in window (b)
for the Au+Au reactions with $b$=10 fm can be similarly understood. We
notice that $R_x^{ME}=(v_0/c)^2$ at the beginning of the collision
is satisfied at all beam energies. As the incident energy increases,
the Lorentz force becomes closer to the Coulomb force.

We now turn to the ratio between the $x$-components of the
nuclear and Lorentz forces, i.e., $R_x^{NM}=F_x^N/F_x^M$, for a
test-proton at the center of mass with a constant velocity of
$v_z=v_{0}$. Shown in windows (c) and (d) are the impact parameter
and beam energy dependences of the $R_x^{NM}$. Because the nuclear
force is proportional to the gradient of the single-nucleon
potential, i.e., $F^{M}=-\nabla_r U$, large fluctuations are seen in
the $R_x^{NM}$. It is seen that the nuclear force is several 10 to
10$^2$ times larger than the Lorentz force. The magnetic field is
thus not expected to affect the reaction dynamics and nucleon
observables. Therefore, it is not surprising that nuclear reaction
models can describe most experimental data without considering any
magnetic effect at all.

\subsubsection{Magnetic effects on collective observables of nucleons and pions}
\begin{figure}[htpb]
\includegraphics[width=1\textwidth]{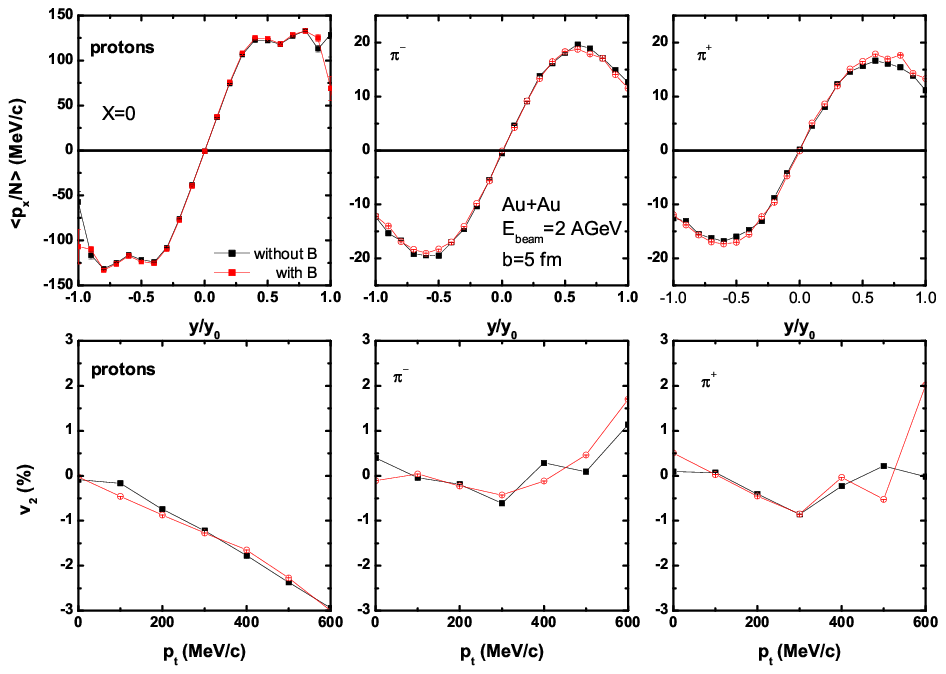}
    \caption{(Top pane) The average in-plane transverse momentum of free protons
    and pions as a function of rapidity.
    (Bottom pane) Elliptic flow for free protons and pions as
    a function of transverse momentum for the 2 AGeV Au+Au reaction
    at an impact parameter of 5 fm with $x$=0. } \label{fv1v2}
\end{figure}
While the magnetic effects on nucleon observables are expected to be
very small,  to be quantitative it is still necessary to examine how
small the effects are. From the expression of the Lorentz force
$F^{M}= q \bm{v} \times \bm{B}$,  it is easy to see that the main
component of the Lorentz force is in the reaction plane (especially
in the $x$-direction). The average transverse momentum in the
reaction plane, i.e., $<p_x>$, is thus a good candidate. Shown in
the top panels of Fig. \ref{fv1v2} are the average in-plane
transverse momentum as a function of rapidity, the so-called
in-plane transverse flow \cite{Danielewicz85}, for free protons and
pions, respectively. Indeed, there is essentially no magnetic effect
on nucleons. It is seen that both negative and positive pions flow
in the same direction as nucleons but with much lower transverse
momentum in the reaction plane \cite{Li94}. Interestingly, there is
a very weak indication of some magnetic effects on the  $<p_x(y)>$ of
pions at forward/backward rapidities. This is qualitatively
understandable because the Lorentz force influences pions motion
easily as they are light compared to nucleons. Moreover, it also
indicates that the magnetic field decreases (increases) very
slightly the magnitude of $<p_x>$ for positive (negative) pions at
both forward and backward rapidities due to the magnetic
focusing/diverging effects as we shall discuss in detail in the next
subsection. Next, we investigate in the lower panels of Fig.
\ref{fv1v2} the so-called differential elliptic flow as a function
of transverse momentum \cite{Ollitrault98,Poskanzer98},
\begin{eqnarray}
\left<v_2(p_t)\right>=\frac{1}{N}\sum_{i=1}^{N}\frac{p_{ix}^2-p_{iy}^2}{p_{ix}^2+p_{iy}^2}
\end{eqnarray}
where $N$ is the total number of free particles. The $p_{iy}$ is
$i$th particle's transverse momentum perpendicular to the reaction
plane. Again, there is essentially no magnetic effect on the differential elliptical flow of both nucleons and
pions.

\subsubsection{Magnetic effects on the $\pi^-/\pi^+$ and neuteron/proton ratio}
\begin{figure}[htpb]
\includegraphics[width=0.5\textwidth]{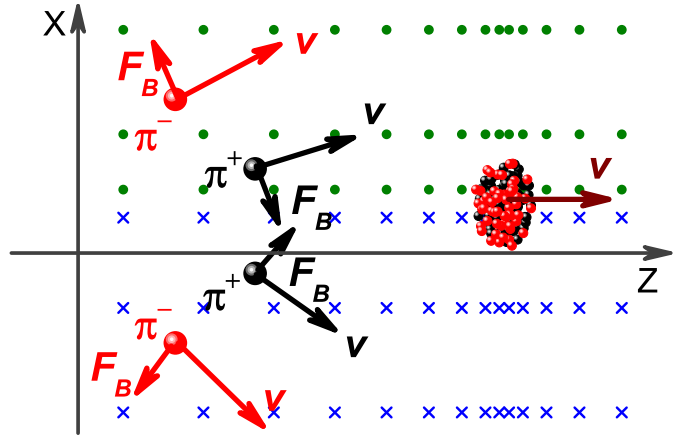}
    \caption{A sketch of magnetic effect on the $\pi^-/\pi^+$ ratio near the projectile rapidity.}
    \label{Bfocus}
\end{figure}

It is well known that the Coulomb force affects significantly the
$\pi^-/\pi^+$ ratio in HICs. The so-called Coulomb peak often
appears near the projectile and/or target rapidities. This
phenomenon has been studied extensively both experimentally
\cite{wolf79,chiba79,benenson,radi82,schn82} and theoretically
\cite{koonin79,bertsch80,gyulassy81,bonasera,nore88,li94} since the
1970's, see. e.g., ref. \cite{stock86} for a review. However,
magnetic effects were not considered in any of these studies. While
the Lorentz force on pions is normally smaller than the Coulomb
force, they have the same order of magnitude. Moreover, compared to
nucleons pions are light with relatively higher speeds and are thus
more easily affected by the Lorentz force. Furthermore, there is no
nuclear force acting on pions once they are produced at least in
most model simulations where pions change their momenta only through
pion-hadron collisions and the Coulomb field. To our best knowledge,
theoretical studies on the mean-field (in-medium dispersion relation) for
pions are still rather inconclusive \cite{Ko10}. Considering all of
the above, the magnetic force on pions can be significant. In fact,
we expect the Lorentz and Coulomb forces to have the opposite
effects on the $\pi^-/\pi^+$ ratio. Namely, near the
projectile/targer rapidity the Coulomb force increase the
$\pi^-/\pi^+$ ratio while the Lorentz force reduces it. Effects of
the Lorentz forces on positive and negative pions are illustrated in
Fig. \ref{Bfocus} using the projectile-like spectator as an example.
The moving track of the spectator can be regarded as a current.
Above/below the current, the magnetic field is perpendicular to the
reaction plane and points outward/inward. The Lorentz force focuses
the $\pi^{+}$ into smaller forward (backward) polar angles while
disperses the $\pi^{-}$ to larger forward (backward) polar angles.
So the $\pi^{-}/\pi^{+}$ ratios at large rapidities are reduced by
the Lorentz force. Moreover, due to the magnetic focusing/dispersing
effect on positive/negative charges, the changes in transverse
momentum for particles above and below the current are opposite. So
the total magnetic effect on the average transverse momentum in the
reaction plane is very tiny even for pions. This explains why the
magnetic effects on the transverse flow $<p_x(y)>$ and the
differential elliptical flow $v_2(p_t)$ are negligible for both
nucleons and pions.

\begin{figure}[htpb]
\includegraphics[width=0.7\textwidth]{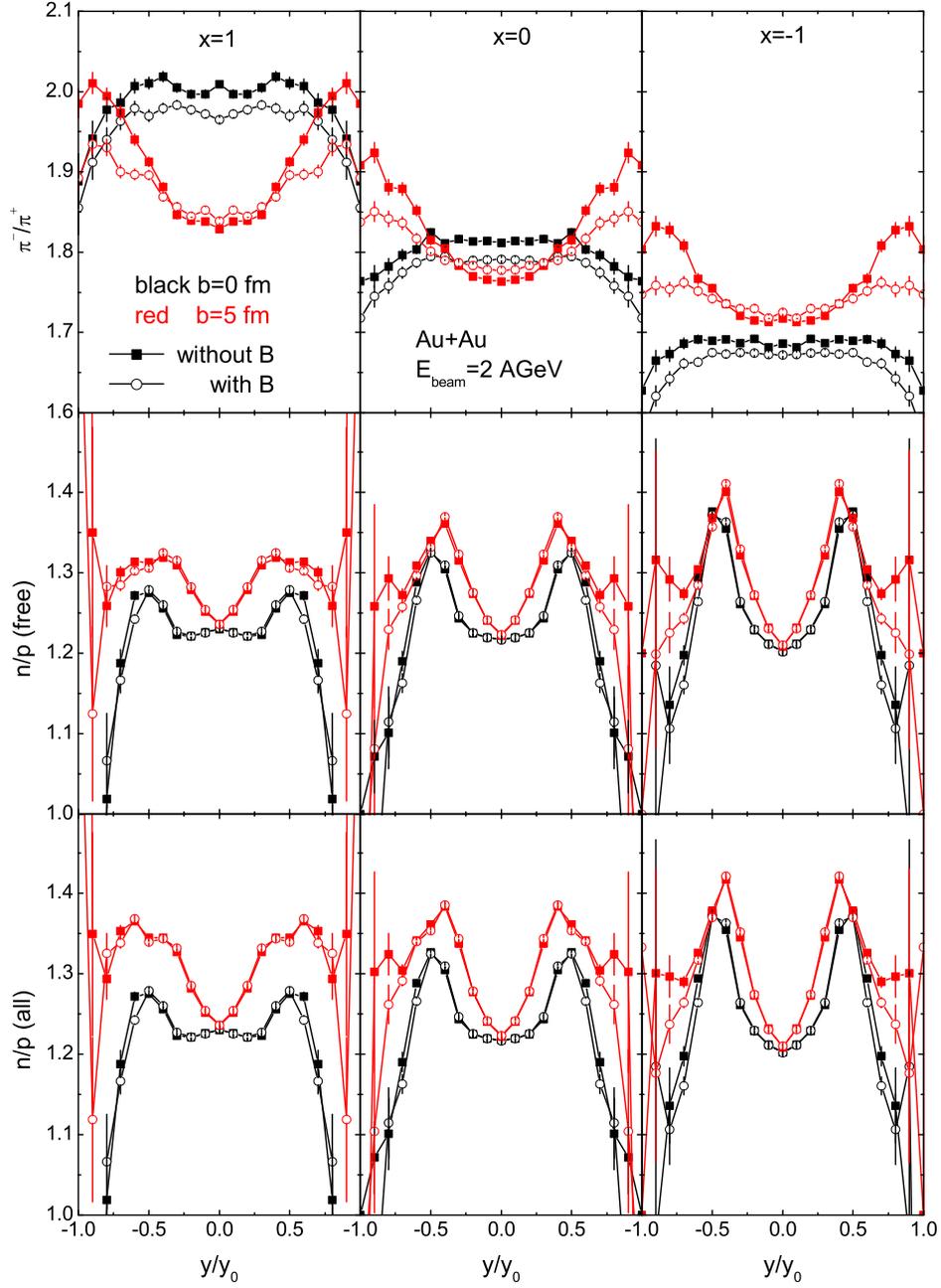}
    \caption{The $\pi^{-}/\pi^{+}$ ratio (top windows),
    the neutron/proton $n/p$ ratio of free (middle windows)
    and all (bottom windows) nucleons as a function of rapidity
    with and without the magnetic field calculated
    with the three different values of symmetry energy parameter $x$
    for the reactions of 2 AGeV Au+Au
    at an impact parameter b=0 and 5 fm, respectively.} \label{ry}
\end{figure}

Why is it so important to understand clearly and precisely the
electromagnetic effects on the $\pi^-/\pi^+$ ratio? One special
reason is that the $\pi^{-}/\pi^{+}$ ratio has been predicted as one
of the most promising probes of the nuclear symmetry energy at
supra-saturation densities \cite{LiBA02}. While comparisons of
transport model predictions \cite{xia,Feng10,Pra} with exiting data
\cite{Rei07} are still inconclusive, all models have consistently
shown that the $\pi^-/\pi^+$ ratio is rather sensitive to the high
density behavior of the nuclear symmetry energy. The latter is
rather poorly known as indicated in Fig. \ref{SymE}. In fact, even
the trend of the symmetry energy at supra-saturation densities,
namely, whether it increases or decreases with increasing density,
is still controversial partially because of our poor knowledge about
the isospin-dependence of strong interaction. To extract reliably
accurate information from the $\pi^-/\pi^+$ ratio about the high-density
symmetry energy, it is thus necessary to understand precisely
effects from the well-known electromagnetic interactions. So, how
strong is the magnetic effect on the $\pi^{-}/\pi^{+}$ ratio in
comparison to the symmetry energy effect? To answer this question
and give a quantitative example, we show in Fig. \ref{ry} the
$\pi^{-}/\pi^{+}$ ratio as a function of rapidity with and without
the magnetic field calculated with three different values of the
symmetry energy parameter $x$ for the 2 AGeV Au+Au reactions at an
impact parameter of b=0 and 5 fm, respectively. In each case
considered here, 200,000 IBUU11 events are used.
Comparing the results obtained with and without the magnetic field
using any of the $x$ parameter considered, it is seen that
significant magnetic effects on the $\pi^{-}/\pi^{+}$ ratio are
obvious especially at forward and backward rapidities particularly
for mid-central collisions. Quantitatively, the $\pi^-/\pi^+$ ratio
obtained with the magnetic field is significantly lower at forward
and backward rapidities (polar angles) due to the magnetic
focusing/diverging effects on the positive/negative pions as we
illustrated in Fig. ~\ref{Bfocus}. Pions at higher rapidities have
larger longitudinal momenta and thus feel stronger Lorentz forces
compared to those at mid-rapidity. For the head-on collisions, the
magnetic effect is small but still appreciable especially in the
early phase of the reactions when most of the pions are produced.
From peripheral to head-on collisions, the $\pi^-/\pi^+$ ratio
changes gradually from forward-backward peaked to center-peaked
distributions. In peripheral collisions, there are significant
Coulomb effects due to the spectators. One thus expects the
$\pi^-/\pi^+$ ratio to peak at forward-backward rapidities. It is
seen that the magnetic effect at forward-backward rapidities is
compatible with the symmetry energy effect from changing the $x$ parameter by one unit. Overall, the
$\pi^-/\pi^+$ ratio decreases as the symmetry energy at
supra-saturation densities becomes stiffer when the parameter $x$
changes from 1 to -1.

It is worth noticing that so far only the
integrated $\pi^{-}/\pi^{+}$ ratio, i.e., the ratio of total $\pi^-$
to $\pi^+$ multiplicities, has been used in attempts to constrain
the symmetry energy at high densities without considering the
magnetic effects. While the integrated $\pi^{-}/\pi^{+}$ ratio is rather sensitive to
the symmetry energy parameter $x$ in reactions near the pion production threshold,
as the beam energy becomes higher than about 1 Gev/nucleon, the sensitivity gradually disappears \cite{xia}.
It is thus interesting to see that the rapidity distribution of the $\pi^{-}/\pi^{+}$ ratio
shows a strong sensitivity to the parameter $x$ even in the reactions at a beam energy of 2 GeV/nucleon where the baryon
density can reach about $3.5\rho_0$. Since the strongest sensitivity to the symmetry energy is at
forward and backward rapidities where the $\pi^{-}/\pi^{+}$ ratio is also
strongly affected by the magnetic field, special cares have to be taken in both model calculations and the data analysis.
Most of the available detectors including the one used by the
FOPI Collaboration \cite{Rei07} do not provide full coverage at very
forward/backward angles. The integrated $\pi^{-}/\pi^{+}$ ratio is
normally obtained by extrapolating the angular distributions of pions measured in a limited angular range to all polar angles. By doing so, however,
the magnetic effects on the angular distribution were neglected.
Previous conclusions on the high-density symmetry energy based on
comparing various transport model calculations with the experimental
data without considering the magnetic effects thus need to be taken
with caution. For comparisons, the neutron/proton ratio $n/p$ of
free (selected as those with local density less than $\rho_0/8$ at
freeze-out) and all nucleons are shown as functions of rapidity
in the middle and bottom windows of Fig. \ref{ry}, respectively. It is seen that
there is essentially no noticeable magnetic effects within error
bars on the $n/p$ ratios. This is consistent with our expectation
and the results on the transverse and elliptical flows discussed
earlier. The non-uniform $n/p$ and $\pi^{-}/\pi^{+}$ ratios as
functions of rapidity indicates the lack of complete isospin
equilibrium for both the nucleon and pion components. This is the so-called
isospin translucency expected in heavy-ion reactions at the beam energies studied here
\cite{LiBA95}.

\begin{table}[!h]
\tabcolsep 0pt \caption{\label{table1} Integrated $\pi^{-}/\pi^{+}$
and $n/p$ ratio calculated without/with the magnetic field using
three values of the symmetry energy parameter $x=1, 0$ and $-1$.}
 \vspace*{-15pt}
\vskip 2mm
 \centering
\def\temptablewidth{1\textwidth}
{\rule{\temptablewidth}{1pt}}
\begin{tabular*}
{\temptablewidth}{@{\extracolsep{\fill}}*{5}{c}
 }
Ratio &$b$ (fm) &$x$=1 &$x$=0&$x$=-1\\
\hline
 \multirow{2}{1.8cm}{$\pi^{-}/\pi^{+}$}
 & 0 & 2.02/1.97& 1.81/1.78& 1.68/1.67\\
 &5 & 1.87/1.86& 1.79/1.79 &  1.73/1.73\\
\hline
 \multirow{2}{1.8cm}{$n/p$ (free)}
 &0 &1.23/1.23 &1.24/1.24 &1.25/1.25\\
 &5 &1.28/1.28 &1.29/1.29 &1.29/1.29\\
 \hline
 \multirow{2}{1.8cm}{$n/p$ (all)}
 &0 &1.23/1.23 &1.24/1.24 &1.25/1.25\\
 &5 &1.31/1.31 &1.31/1.31 &1.32/1.32\\
\end{tabular*}
{\rule{\temptablewidth}{1pt}}
\end{table}
Shown in Table \ref{table1} are the integrated $\pi^{-}/\pi^{+}$ and neutron/proton
ratios calculated without/with the magnetic field. It is seen that the
integrated ratios are not affected much by the magnetic field. This
is what we expected as the Lorentz force affects differently only
the angular distributions of positively and negatively charged
particles, but not their total multiplicities. Also, consistent with
previous findings \cite{xia} the integrated $\pi^{-}/\pi^{+}$ ratio
at beam energies higher than about 1 GeV/nucleon is not so sensitive
to the variation of the symmetry energy while there is a clear
indication that a higher $\pi^{-}/\pi^{+}$ ratio is obtained with a
softer $E_{\rm{sym}}(\rho)$ at supra-saturation densities. Thus, the
differential $\pi^{-}/\pi^{+}$ ratio as a function of rapidity, as we discussed earlier, is a
better probe of the symmetry energy at supra-saturation densities
after taking care of the magnetic effects.

\section{Summary}
In summary, within the transport model IBUU11, the time-evolution
and space-distribution of internal electromagnetic fields in HICs at
beam energies between 200 and 2000 MeV/nucleon are studied. While
the magnetic field can reach about $7\times 10^{16}$ G, it has
almost no effect on nucleon observables as the Lorentz force is
normally much weaker than the nuclear force. On the other hand, the
magnetic field has a strong focusing/diverging effect on
positive/negative pions at forward/backward rapidities.
Consequently, the differential $\pi^-/\pi^+$ ratio as a function of
rapidity, but not the integrated one, is significantly altered by
the magnetic field. At beam energies above about 1 GeV/nucleon, the
differential $\pi^-/\pi^+$ ratio is more sensitive to the
$E_{\rm{sym}}(\rho)$ than the integrated $\pi^-/\pi^+$ ratio. Our
findings suggest that magnetic effects should be carefully
considered in future studies of using the differential $\pi^-/\pi^+$
ratio as a probe of the $E_{\rm{sym}}(\rho)$ at supra-saturation
densities.

\section{Acknowledgements}
We would like to thank Drs. N. Chamel, W. G. Newton and C. Providencia for
helpful discussions and information on magnetic effects in neutron
stars, Dr. Lie-Wen Chen and Dr. Chang Xu for collaborations in
developing the IBUU11 code used in this study. We would also like to
thank Dr. Derek Harter who made our very intensive calculations possible
within a rather short time by providing us access to the high-performance
Computational Science Research Cluster at Texas A\&M University-Commerce.
This work was supported in part by the NSF under grants PHY-0757839 and PHY-1068022 and NASA
under grant NNX11AC41G issued through the Science Mission
Directorate, and the National Natural Science Foundation of China
under Grant Nos 11005022, 10847004 and 11075215.


\end{document}